# AR Based Half-Duplex Attack in Beyond 5G networks

Misbah Shafi, *Student Member, IEEE,* Rakesh Kumar Jha, *Senior Member, IEEE,* and Manish Sabraj

*Abstract*—With the evolution of WCN (Wireless communication networks), the absolute fulfillment of security occupies the fundamental concern. In view of security, we have identified another research direction based on the attenuation impact of rain in WCN. An approach is initiated by an eavesdropper in which a secure communication environment is degraded by generating Artificial Rain (AR), which creates an abatement in the secrecy rate, and the cybersecurity gets compromised. By doing so, an attacking scenario is perceived, in which an intruder models a Half-Duplex (HD) attack. Half-Duplex specifies the attack on the downlink instead of targeting both uplink and downlink. This allows the attacker to alleviate the miss-rate of the attacking attempts. The layout for the HD attack is explained using RRC (Radio Resource Control)-setup. Further, we have determined and examined the performance parameters such as secrecy rate, energy efficiency, miss-rate, sensitivity in the presence of AR. Further comparison of rural and urban scenarios in the presence and absence of AR is carried out concerning the variation in secrecy rate with respect to the millimeter-wave frequencies and distance. Lastly, the methodology of the HD attack is simulated, revealing that the HD attack maintains a low miss rate with improved performance as compared to the performance and miss-rate attained by the full-duplex attack.

*Index Terms*— Artificial Rain (AR), HD attack, miss-rate, D2D (Device to Device) communication, secrecy rate, sensitivity.

## I. INTRODUCTION

The visualization of 6G (Sixth Generation) reflects the presumption of the user demand, onward novelties, technological potentials in the form of ultra-high data rates (100-1000 Gb/s), long-range services, ultralow-power usage, low latency [1]. Since 5G is expected to justify 10x more connectivity density [2], 100x more energy efficiency [3] and, 1000x more capacity [4] as compared to the existing communication network. With the progression of such requirements, certain advances were encompassed in 5G, such as the involvement of clouds in network communication and processing, flat architectures, and heterogeneous networks (WiFi hotspots, microcells, small cells, femtocells) [5]. Thus, due to centralized architecture, the susceptibility of security threats increased [6]. From preceding years, security threats involve numerous attacks on the measures of cybersecurity, such as confidentiality, integrity, privacy, and authentication.

### A. Related work

The paramount phenomenon of the weather is the occurrence of the hydrometeors. The existence of the hydrometeors occurs in the form of rain, graupel, hail, and snow produce the effect of absorption and scattering. This effect creates an attenuation impact on the propagation of electromagnetic waves. These are severely affected at frequencies greater than 10 GHz due to the reduction of power density [8]. In case of 5G NR (New Radio), the second band of frequency 24 GHz to 52.6 GHz [17] incorporates higher frequencies. The particles of the rain in the wireless channel at higher frequencies create the stern impact of scattering and absorption in the path of the transmitted signal [18], [19]. In the present era of WCN, the attenuation due to rainfall on the propagating signal is not equitably approximated. The attenuation of the signal has a considerable effect when the diameter of the raindrop lies in the range from 0.1mm to 10mm is less than the operating wavelength of the signal. An impact of rainfall attenuation in mmWave (Millimeter Wave) MIMO system shows an increase of capacities and achievable rates at low rain rates due to a decrease of antenna correlation. At medium and high rain rates, there occurs a decrease of capacity and achievable rates with an increase in rain rates [7].

Various approaches were followed to determine the impact of rainfall attenuation on the propagating signal. The approaches involved in these prediction methods include ITU-R P.837-6 (International Telecommunication Union Radio Communication) sector, MORSE (Model for Rainfall Statistics Estimation), and ITU-R P.837-7. These prediction methodologies are compared on the basis of the accuracy of prediction [8]. A global model of prediction provides much higher accuracy and more excellent spatial resolution by making the use of CHIRPS (Climate Hazards Group Infrared Precipitation with Stations) [10]. Another approach involves the performance analysis of mmWaves under the rainfall and non-rainfall scenario. The analysis is determined on account of WPT (Wireless Power transfer) under the consideration of rainfall and non-rainfall effects, small scale fading, path loss, and the energy propagation models [9]. Effective use of calculated signal attenuation in the presence or absence of rainfall can be utilized to determine the accumulation of hydrometeors along with the communicating link. Satellite link signal attenuation operating at 12.3 GHz can be exploited to estimate the accumulation of rainfall in the path link. To compute such, the power-law relation between the rate of attenuation and rate of rainfall is evaluated [11].

D2D communication executes a crucial role in next-generation WCNs. It possesses the ability to offload massive data traffic by enlightening the utilization of network resources, providing short-range communication and proximity-based services . In spite of its numerous advantages, several security

TABLE I
COMPARISON OF THE PROPOSED SCHEME WITH THE EXISTING SCHEME

| Reference | Objective | Associate parameters | General Analysis | Security Impact | Scenario covered |
|---|---|---|---|---|---|
| [7] | Impact of rain on achievable capacities of the mm-Wave MIMO system | Rain attenuation, bit rate, capacity | ✓ | ✗ | Rain |
| [8] | To estimate the accuracy of rainfall rate prediction methods | Prediction accuracy, rain rate, error figure | ✓ | ✗ | Rain |
| [9] | Evaluation of the performance of WPT in the non-rainy and rainy condition of an mm wave MIMO system | Normalized harvested energy, retrieval factor, area of circular coverage, average power | ✓ | ✗ | Rain |
| [11] | To analyze the rainfall accumulation on the communicating link using attenuation of satellite signal | Rainfall accumulation, length of the link, effective rain height | ✓ | ✗ | Rain |
| [13] | Single and multi-channel downlink resource management for the improvement of PLS | D2D sum rate, cellular user (CU) sum rate, CU security probability | ✓ | ✓ | D2D |
| [14] | To improve the PLS using an access selection scheme in a cellular D2D communication network. | Secrecy throughput, optimal threshold, the density of eavesdroppers, outage probability | ✓ | ✓ | D2D |
| [15] | A security embedded interference avoidance scheme is evaluated to improve security by decreasing interference. | Symbol error probability, SNR | ✓ | ✓ | D2D |
| Proposed | The security impact of AR followed by Half-Duplex attack | Miss-rate, rain attenuation, secrecy rate, sensitivity, energy efficiency | ✓ | ✓ | AR, Half-Duplex Attack in D2D communication |

*AR=Artificial Rain, PLS=Physical Layer Security

menaces are emerging with an increasing number of devices [12]. Another scheme of security improvement includes the combined optimization of channel assignment and power allocation of communicating D2D links where downlink resource sharing is used via multi-channel and single channel communication [13]. For multiple eavesdroppers, one of the approach to enable secure communication is the access selection scheme, where D2D users are present at the distance greater than the threshold such that interference is created by the specific D2D user to craft the jamming impact on the attacker [14]. A security embedded interference avoidance scheme is evaluated where the constellation rotation technique is used. Thereby improving the PLS (Physical Layer Security) of the network [15]. Another mechanism that can improve the security of the relay device is SLNR (Signal to leakage and noise ratio) precoding scheme. It improves the security of the network by balancing out the noise and interference [27].

*B. Contribution*

In this paper, the attack modelling is proposed in next generation WCN. The target scenario inspects the D2D communication network. The attack involves the execution of AR followed by the commencement of HD attack. The fundamental contribution of this paper is specified as follows:

1. A system model is signified with an unknown CSI (Channel State Information). The intruder is capable of monitoring the location of the users from the BS (Base Station). The user with worst channel conditions is targeted for the attack.
2. The AR attenuation incorporating security analysis is implemented on the user with deteriorated CSI. The examination is performed in rural and urban scenarios to define the secrecy rate analysis for the attack.
3. HD attack is initiated on the targeted user. The resources that are supposed to be allocated to the valid user are spoofed by the intruder.
4. The comparison on the basis of performance in terms of sensitivity, miss-rate, and energy efficiency is evaluated between HD, and FD (Full-Duplex) attack.

The organization of the paper is mentioned as: The system model and the problem formulation is represented in Section II. Section III demonstrates the general mechanism of the HD attack. An Illustrative Mechanism of HD attack with the help of an example is discussed in Section IV. The analysis of the simulation results is performed in Section V. Finally, the conclusion of the paper is specified in Section VI

II. SYSTEM MODEL AND PROBLEM FORMULATION

The system model demonstrating the scenario of AR, followed by the problem formulation, is investigated in this section.

*A. System Model*

The proposed system model is characterized by two case scenarios, followed by the execution of an HD attack. The first case determines the background of WCN in the absence of an eavesdropper without AR devoid of an intruder. The recent mechanism for the generation of AR is given in [20]. The second case demonstrates the framework of WCN in the presence of an intruder, triggering AR. Further HD attack is formulated in the communication scenario incorporating security analysis.

*1) Case I: Scenario of WCN devoid of an intruder*

Consider a signal under 5G WCN in which the transmitter $A$ at a distance of $r$ transmits a signal $y(t)$ to receiver $B$ or vice versa. Assume negligible, or no hindrance condition in the communicating link in reference to the transmitter and receiver representing free space WCN and the signal is propagating in a straight path. The path loss of free space presents a complex factor of scaling in the receiving signal such that the received signal $z(t)$ can be calculated as:

$$z(t) = Re\left\{\frac{\lambda\sqrt{B_l}\,e^{-\frac{j2\pi r}{\lambda}}}{4\pi r} y(t)\, e^{-2\pi f_c t}\right\}, \quad (1)$$

where $B_l$ is the product of field radiation patterns of transmit antenna and receive antenna, $e^{-\frac{j2\pi r}{\lambda}}$ denotes the phase shift that arises due to the traveling distance $r$ of the wave from transmit antenna to receive antenna, $\lambda$ depicts the wavelength of radiation, $y(t)$ denotes the complex baseband signal and is the combination of an in-phase component, and quadrature component is given by:

$$y(t) = u_I(t) + ju_Q(t) \quad (2)$$

where $u_I(t)$ is the real part of $y(t)$ and denotes in-phase component and $u_Q(t)$ is the imaginary part of $y(t)$ and denotes the quadrature component. The ratio of the received power $P_r$ to transmit power $P_t$ is given by:

$$\frac{P_r}{P_t} = \left[\frac{\sqrt{B_l}\lambda}{4\pi r}\right]^2 \quad (3)$$

Approximated path loss model as a function of distance $r$ is expressed as:

$$P_r = P_t\, q \left[\frac{r_o}{r}\right]^{\psi} \quad (4)$$

where $q$ is a dimensionless constant depending on the characteristics of average attenuation of channel and antenna, $r_o$ is the distance of reference, $\psi$ is the path loss exponent. Where $q$ can be expressed as:

$$q_{(db)} = 20 \log_{10}\left(\frac{\lambda}{4\pi r_o}\right) \quad (5)$$

Using a model of log-normal shadowing, suppose the ratio of transmit power to receive power is represented by '$\phi$' be random, given by log-normal distribution as:

$$p(\phi) = \frac{\Delta}{\sqrt{2\pi}\,\sigma_{\phi\,(db)}\,\phi} \exp\left\{-\frac{(10\log_{10}\phi - \mu_{\phi\,(db)})^2}{2\,\sigma^2_{\phi\,(db)}}\right\}, \phi > 0 \quad (6)$$

where $\Delta$ is a constant and is given as:

$$\Delta = \frac{10}{\ln 10} \quad (7)$$

where $\mu_{\phi\,(db)}$ denotes the mean of $\phi_{(db)}$, $\sigma^2_{\phi\,(db)}$ represents the variance of $\phi_{(db)}$ and $\sigma_{\phi\,(db)}$ represents the standard deviation of $\phi_{(db)}$. The expectation of $\phi$ or the linear average path gain or the mean of $\phi$ can be given as:

$$E[\phi] = \mu_\phi = exp\left[\frac{\mu_{\phi\,(db)}}{\Delta} + \frac{\sigma^2_{\phi\,(db)}}{2\Delta^2}\right] \quad (8)$$

Equation (6) can also be written as:

$$p(\phi_{(db)}) = \frac{1}{\sqrt{2\pi}\,\sigma_{\phi\,(db)}} \exp\left\{-\frac{(\phi_{(db)} - \mu_{\phi\,(db)})^2}{2\,\sigma^2_{\phi\,(db)}}\right\} \quad (9)$$

The combined path loss and equation (9), the shadowing can be evaluated as:

$$P_{l\,(db)} = -10 \log_{10} q + 10\psi \log_{10}\frac{r}{r_o} + \phi_{(db)} + e^t \quad (10)$$

where $e^t$ denotes the thermal energy component. The SNR can be evaluated as:

$$\varsigma = \frac{|K|^2 P_t}{\gamma'} \quad (11)$$

where $|K|^2 = -P_{l\,(db)}$ and defines the channel gain. The capacity denoted by $C_u$ can also be expressed using the theorem of Shannon's capacity as:

$$C_u = \log_2(1 + \varsigma) \quad (12)$$

Where $C_u$ is the capacity of a legitimate user without eavesdropper's intervention.

*2) Case 2: Attacking scenario using Artificial Rain (AR)*

Consider the WCN in the presence of an intruder. An assumption is made that the intruder is capable of achieving the location of the users from the Base Station (BS). The location of the users from the BS is achieved by TPA [22] technique. As shown in Fig. 2, in the region 1, the users distant from the BS require more energy and are thereby less energy efficient.

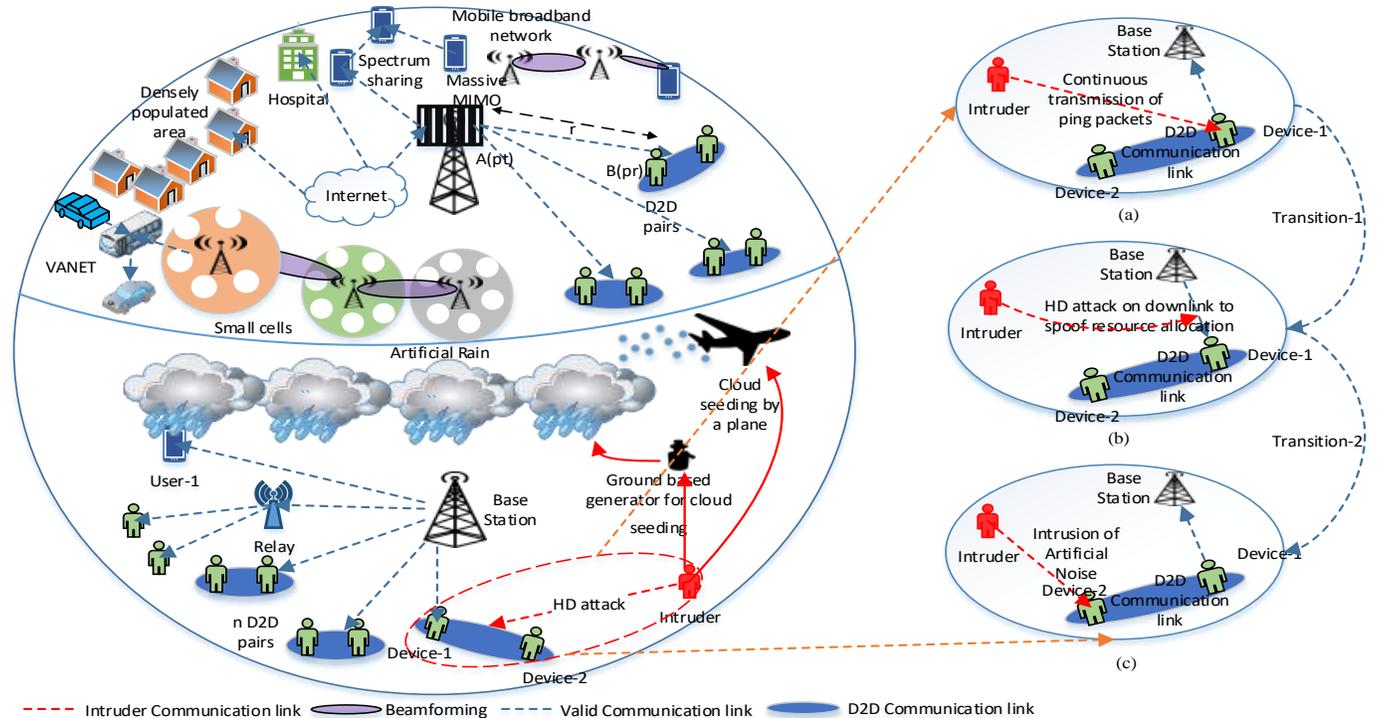

Fig. 1. System model 1(a) Process of ping packet transmission 1(b) Process of resource spoofing1(c) Process of AN intrusion

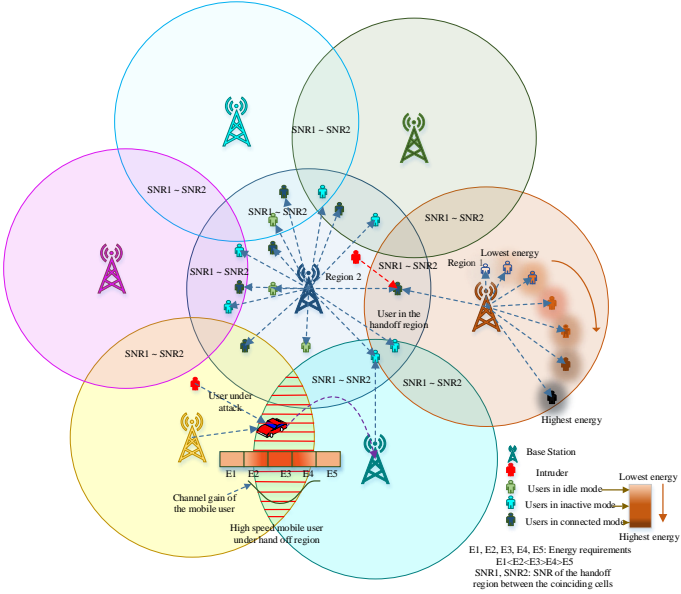

Fig. 2. Location tracking of the user using TPA technique

In other word, with an increase in distance from the BS, the energy requirement tends to increase. Therefore, following the footprint of energy pattern analysis the location of the users is determined. Based on the parameter of distance, CSI is analyzed by using the mathematical modelling given in this section. To degrade the SNR of the communication channel, an attempt is made by an intruder in the form of artificial seeding of the clouds to create AR. The attenuation occurring in a signal while traveling through obstacles in the form of rain droplets with depth $d_R$ is given by:

$$d_R = d_{SC} + d_{AB} + d_{POL} \quad (13)$$

Such that attenuation due to rainfall can be given as:

$$x'(d_R) = e^{N_R d_R} \quad (14)$$

Where $x'(d_R)$ denotes the attenuation of a signal due to artificial rain. The equation (10) for path loss, including the effect of artificial rain attenuation, can be expressed as:

$$P'_{l\,(db)} = -10 \log_{10} q + 10\psi \log_{10} \frac{r}{r_o} + \phi_{(db)} + \left(x'(d_R)\right)_{db} + e^t \quad (15)$$

TABLE II
TABLE OF SYMBOLS

| Symbol | Description |
|---|---|
| $\lvert K\rvert^2, \lvert K'\rvert^2, K_{ij}$ | Path gain, path gain in the presence of AR, channel coefficient from $ith$ node to $jth$ node and vice versa. $i = A, B$ and $j = A, B$, where $i \neq j$ |
| $y(t), y_i,$ | Transmitted signal, transmitted signal by the corresponding device $i$ |
| $z(t), z_i$ | Received signal, received signal by the corresponding device |
| $p_l, p'_l$ | Path loss in free space, in the presence of AR |
| $\varsigma, \varsigma'$ | Signal to noise ratio in the absence of intruder, in the presence of AR |
| $C_s(C_s)_{AR}, C_T, C_u, C_{ev}, (C_{ev})_{AR}$ | Secrecy capacity of a valid user, secrecy capacity in the presence of AR, threshold capacity, capacity of the valid user, capacity of the eavesdropper introducing AR |
| $d_{SC}, d_{AB}, d_{POL}$ | Depth of scattering, absorption, and polarization in raindrop |
| $\gamma', N_{ev}, n_t$ | Noise power, Artificial noise introduced by the intruder, AWGN at corresponding time interval $t$ |

$N_R$ is the specific attenuation and depends on the dielectric properties of the rain droplet and its composition. $N_R$ due to AR can be determined by using power-law relationship given as:

$$N_R = \theta R^\varepsilon \quad (16)$$

Where $\theta$ and $\varepsilon$ are the coefficients of specific attenuation and depend on the frequency $f$ (in GHz). These coefficients are expressed as:

$$\log_{10} \theta = \sum_{i=1}^{4}\left(\delta_i \exp\left[-\left(\frac{\log_{10} f - \zeta_i}{\vartheta_i}\right)^2\right]\right) + a_k \log_{10} f + b_k \quad (17)$$

where $f$ is the frequency, $a, b, \delta, x,$ and $y$ are coefficients of $\theta$, and $\varepsilon$ can be determined as :

$$\varepsilon = \sum_{i=1}^{4}\left(\delta_i \exp\left[-\left(\frac{\log_{10} f - x_i}{y_i}\right)^2\right]\right) + a_k \log_{10} f + b_k \quad (18)$$

The coefficient $\theta$ can be determined either for horizontal polarization $\theta_H$ alternatively, for vertical polarization $\theta_V$. Similarly, $\varepsilon$ can also be determined for horizontal polarization ($\varepsilon_H$) or vertical polarization ($\varepsilon_V$). For circular, linear and other path geometries the coefficients $\theta$ and $\varepsilon$ can be calculated as:

$$\theta = \frac{(\theta_H + \theta_V + (\theta_H - \theta_V)\cos^2\alpha \cos 2\beta)}{2} \quad (19)$$

$$\varepsilon = \frac{(\theta_H \varepsilon_H + \theta_V \varepsilon_V + (\theta_H \varepsilon_H - \theta_V \varepsilon_V)\cos^2\alpha \cos 2\beta)}{2\theta} \quad (20)$$

Where $\alpha$ is the angle of path elevation and $\beta$ is the tilt angle of polarization relative to the horizontal. Also, channel gain with the effect of AR can be given as:

$$\lvert K'\rvert^2 = -P'_{l\,(db)} \quad (21)$$

Therefore, SNR and capacity can be given as:

$$\varsigma' = \frac{\lvert K'\rvert^2 P_t}{\gamma'} \quad (22)$$

$$(C_s)_{AR} = \log_2(1 + \varsigma') \quad (23)$$

### B. Analysis of secrecy capacity

The approach to analyze the security of the communication network involves the parameter of secrecy capacity, obtained from wireless wiretap theory. It is defined as the difference between channel capacities of valid user and intruder, given by:

$$C_s = [C_u - (C_{ev})]^+ \quad \forall \ C_u \geq C_{ev} \quad (24)$$

For a secured WCN

$$C_s > C_T \quad (25)$$

The intruder attempts to decrease the secrecy capacity below threshold capacity by creating the attacking scenario such that

$$C_s < C_T \quad (26)$$

To achieve such, AR is introduced by the intruder in the communication setup where attenuation due to AR decreases the secrecy capacity as compared to the secrecy capacity achieved without intruder given by:

$$C_s > (C_s)_{AR} \quad (27)$$

The total of secrecy capacity is required to be reduced to satisfy the criteria of equation (26) for the initiation of an attack. It depends upon the sum of rain rate added in the communicating environment. However, the required amount of AR attenuation

is identified by the intruder such that $C_s$ reduces below $C_T$ given by:
$$(C_s)_{AR} = C_u - (C_{ev})_{AR} < C_T \quad (28)$$
It can be concluded, that with an increase in AR attenuation, the secrecy capacity reduces correspondingly. Therefore, at a specific rate of AR attenuation secrecy rate reduces below the threshold capacity. The prime aspect of the system model shows the effect on the secrecy rate of the channel while degrading the capacity of the main channel. The resultant observation from equation (26) and (28) predicts an increased chance of successful spoofing. **(See Lemma 1)** ∎

***Remarks 1:*** *The threshold capacity is obtained at a threshold distance, and in our scenario, threshold distance is achieved to be 150m. Below threshold distance, the achieved capacity of the valid user is greater than the capacity obtained by the intruder capacity.*

The attack modelling can prove advantageous to strengthen the defence communication network of a nation. The resultant execution can jam the communication network of the opponent. In case of defence communication network, the proposed model can prove cost efficiency in contrary to the cost of other warfare models of the defence. Table III provides a description of various preceding attacks. It provides the idea of ongoing strategies involved in an attack.

### III. GENERALIZED MECHANISM OF HALF-DUPLEX ATTACK

This section illustrates the mechanism of the proposed attack investigating cybersecurity by the processes computed.

#### A. Overview of the proposed attack

Security occupies immense attention in future generation communication technologies, such as spectrum sharing and D2D communication. These technologies involve the self-monitored network and are less secure due to the involvement of relay devices. The FD relays in case of small cells are mostly configured to improve the coverage range and correspondingly the throughput [26]. The proposed attack encompasses D2D communication having device relaying controlled link, assisted by the BS configuration wherein attack is posed on the relay device. The attack targets downlink exclusively in comparison to the FD attacks, aiming at both uplink and downlink. Further, the proposed attack is devoid of the authentication phase, which reduces the chance of miss-rate. Therefore, it is predicted that the proposed attack employs greater efficacy and security threat, comparative to the FD attacks. Consider the scenario wherein $n$ number of D2D pairs are present in the network, and device-2 is present at the edge of the cell where the signal of the BS is not reachable with sufficient strength. The device-1 is present adjacent to the BS and is also approachable by the device-2. Following the device relaying controlled link configuration, device-2 is able to access the communication via device-1, as shown in Fig.1. Assume an Eavesdropper is present neighboring to device-1 and device-2, respectively, observes the communication continuously and has sufficient energy sources to introduce the attack. The intruder is known of the location of the valid users present in the communication network. The location of the users is obtained by the Thermal Pattern Analysis (TPA) technique.

The user farther from the BS requires more energy in contrary to the nearby users. The users are analyzed based on the footprint of TPA [21]. The user with highest energy requirements is correspondingly more distant from the BS. Therefore, more energy is required to transmit the signal from BS to the respective user. In Fig 2. region 1 shows the TPA with respect to the distance of the user from the base station. From the parameter of distance, the path loss is evaluated. Correspondingly the CSI is investigated by the inspection of path loss analysis. The comparison is made between the CSI of each user present in the network. The user with the most degraded CSI is selected for the attack.

#### B. Formulation of the proposed attack

The proposed mechanism of half-duplex attack has the objective of minimizing the miss rate probability $P(m_r)$ and to maximize the effectiveness $A_p$ of the attack shown as:
$$A_p = \frac{1}{\min\{P(m_r)\}} \quad (29)$$
The attack can be launched by evaluating the following processes, as shown in Fig 1.(a), 1(b), 1(c) while considering transmission time intervals for RRC (Radio Resource Control) setup configuration, as shown in Fig. 3.

#### 1) Process of ping packet transmission

In our formulation, this process of an attack aims to make the relay device-1 inaccessible to the BS during the downlink. During the phase of uplink, authentication takes place by device-1 at the BS. It is followed by the continuous transmission of ping packets by an intruder at the relay device-1. Generally, ping packets (request-response) are used to determine the connectivity between the transmitter and the receiver. The continuous transmission of ping packet requests by an attacker to the relay device-1 results with an equal number of ping reply packets along with the identification number and the sequence number from the device to the intruder. The procedure of ping request and ping reply with ID number provides the chance to spoof the identity of the target user. The continuous procedure of ping flood leads to the overload of network connectivity and inefficient drain of resources. This process of attack targets the availability parameter of the security and makes the device-1 unavailable for the downlink of BS. The process of Game Theory [22] scheme can also be incorporated to spoof the ID of the valid user. The process flow of ping packet reception for timestamp $t_n$ is expressed as follows:

$$Z_n = \sum_{i=2}^{n}(K_{ev1} r_i' + n_n) \quad (30)$$

TABLE III
TABLE OF ATTACKS

| Attack type | Description | Target parameter |
|---|---|---|
| Signaling storms | Traffic patterns that overload the control plane | Availability |
| IP spoofing | Creation of IP packets with the forged IP address to conceal source identity | Availability |
| Scanning attack | Open interface attacks | Confidentiality |
| Frame injection | The attacker injected their own frames and crafted the original one. | Integrity |
| Rogue access point | Establishing an unsafe access point inside the firewall. | Confidentiality |
| Radius cracking | Hacking RADIUS authentication server to get login credentials. | Authentication |
| Bandwidth spoofing | Flooding a network to the extent that they start affecting legitimate traffic | Availability |
| Fraggle attack | Sending spoofed UDP packets for broadcast in the network. | Availability |
| Malware attack | The malware is penetrated into the network via software components leads to disruption of ordinary functionality | Availability |

Where $Z_n$ is the summation of $n$ received ping requests by device-1. $n_n$ is the AWGN (Additive White Gaussian Noise) with zero mean and $v^2$ variance $n_n \sim \mathbb{N}(0, v^2)$, $r'$ is the transmitted ping by the intruder, $K_{ev1}$ depicts the channel coefficients from eavesdropper to the device-1. The synchronization of the intruder with the BS and its users is achieved by the mechanism of TPA [21] while examining the BS and its users in idle and active mode. The energy required for the idle mode is comparatively less as compared to that of the active mode. Therefore following the footprints of the thermal energy patterns of the user, the intruder achieves the synchronization. The idle mode and active mode reference signal mechanism in view of energy requirements is defined by ultra lean design [23], [24].

*2) Process of resource spoofing*

The primary purpose of the proposed attack is the spoofing of resources with reduced miss rates. This process involves the successful reception of downlink by the intruder, transmitted by the BS. The procedure is employed by masquerading the ID of device-1 by the intruder such that the resource allocated meant to be transmitted by the BS to the valid device is received by the intruder. Thus, instead of targeting both uplink (authentication phase), solely downlink is targeted. From Fig. 3 at Transmission Time Interval (TTI) $t_1$, RRC setup request is transmitted by the device-1 to gNB for the completion of authentication. The TTI $t_1$ is followed by the $t_2, t_3, t_4, \ldots, t_n$ Where the process of ping flood takes place up to $n$ time intervals, as shown in equations (30). Due to which intruder can conceal its identity with the identity of valid device-1 for $t_{n+1}$ timestamp. At TTI $t_{n+1}$, downlink from gNB, RRC setup response is transmitted to the intruder instead of valid device-1, such that intruder RRC-IDLE obtains the intruder RRC connected state, as shown in Fig. 3. During time stamp $t_{n+2}, t_{n+3}$ the random sequence is transmitted to device-2, and an RRC setup completion acknowledgment is sent to gNB. For the case of a connection failure mechanism, an intruder fails in achieving the connected state. The intruder continuously repeats the process from the TTI $t_2$ until connected RRC state is achieved.

*3) Process of Artificial Noise (AN) intrusion*

The AN is transmitted by the intruder to the device-2 during the transmission of an information-bearing signal from device-1 to device-2 at the time stamp $t_{n+2}$. The intruder acts as a jammer to interfere in the signals between device-1 and device-2. The device-2, consequently, attempts to decode the interfered signal. The purpose of incurring artificial noise at device-2 is to prevent the request signal back to BS from the device-2. Where AN can be a randomly generated sequence. The received signal at the device-2 with artificial noise can be given as:

$$Z_2 = K_{12} y_1 + K_{ev2} N_{ev} + n_{t_{n+2}} \quad (31)$$

The formulated attack aggregates the process over multiple time intervals. For a range of challenging capability of attacks, a HD attack exhibits greater strength to evolve the attack successfully. The analysis of the probability of a successful attack can be explained in the next section.

*C. Impact on miss-rate probability*

In Fig. 3. RRC (Radio Resource Control) setup executing half-duplex scenario is represented. The downlink reception from BS for the timestamp $t_{n+1}$ involves two possibilities. The first possibility includes the successful reception of downlink RRC setup by the legitimate device-1, denotes failure as $m_{DL}$ while as the second possibility involves the successful reception of downlink RRC setup by an intruder, depicting success as $b_{DL}$ The corresponding probabilities of missed attempts and successful attempts define the probability of missed attempts and the probability of successful attempts of the HD attack. Let $Z[t_i]$ for $i = 1,2,3, \ldots n, n+1, n+2, \ldots$ and $n = 1,2,3, \ldots$ defines a Bernoulli process using parameter $m_{DL}$. If $Z[t_1], Z[t_2], Z[t_3], \ldots, Z[t_i]$ are Independent and Identically Distributed (I.I.D.) random variables such that the process is defined as a discrete random process whose density function is expressed as:

$$f_{Z_{t_i}[t_i]}(z_{t_i}[t_i]) = b_{DL} \delta(z_i[t_i]) + m_{DL} \delta(z_{t_i}[t_i] - 1) \quad (32)$$

Where $\delta(.)$ depicts the unit impulse function. At the intruder, for the half-duplex attack, the sum of the Bernoulli processes can be obtained as:

$$S(z_{t_i}[t_i] = u) = z_{t_1}[t_1] + z_{t_2}[t_2] + z_{t_3}[t_3] + \cdots + z_{t_i}[t_i] \quad (33)$$

The probability of successful reception of downlink by the device-1 during a half-duplex attack or the probability of missed attempt can be obtained by using equation (33) such that the sum of the Bernoulli processes defines the binomial process.

$$P(S(z_i[t_i] = u))_{HD} = \binom{i}{u} m_{DL}^u b_{DL}^{i-u}$$
$$= \frac{i!}{(i-u)! u!} m_{DL}^u b_{DL}^{i-u} \quad (34)$$

Similarly successful reception at the intruder is given by:

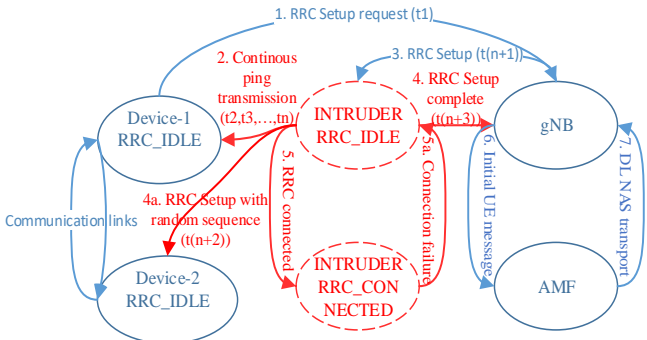

Fig. 3. RRC setup execution under HD attack

$$P_1(S(z_i[t_i] = u))_{DL} = \binom{i}{u} b_{DL}^u \, m_{DL}^{i-u} \quad (35)$$

For FD attack, the involvement of uplink and downlink takes place such that the missed attempts can be evaluated for both uplink and downlink. The probability of missed attempts defines miss rate probability. The corresponding probabilities of missed attempts for uplink and downlink in FD attack is given by:

$$P(S(z_i[t_i] = u))_{FD} = \binom{i}{u} m_{FD}^u \, b_{FD}^{i-u} \quad (36)$$

Let the probability of missed attempts in half-duplex and probability of missed attempts in the downlink of FD attack are equal such that:

$$P(S(z_i[t_i] = u))_{HD} \leq P(S(z_i[t_i] = u))_{FD} \quad (37)$$

The equation (37) indicates that the miss rate probability of accessing resource allocation in a HD attack is less than the miss rate probability achieved in the FD attack. However, the probabilities of missed attempts in HD attack and FD attack can be equal for zero probability of missed attempts during uplink in the FD attack. **(See Lemma 2)** ∎

*Corollary 1:* The miss-rate achieved for the HD attack $(m_r)_{HD}$ on device-1 by an intruder is less than or equal to miss rate of intruder achieved in the FD attack $(m_r)_{FD}$.

$$(m_r)_{HD} \leq (m_r)_{FD} \quad (38)$$

The proof of the corollary 1 is shown in the appendix.

*D. Case of mobile user*

In case of the mobile user, the intruder manages to achieve the impact of AR by executing the AR over a large area. AR lies in the range of 0.5Kms [25]. The next generation communication incorporates small cell radius occupies the same range as of the AR coverage. The procedure of the AR covers the area of the mobile user. In other words, the range of the area in which mobile user roams in a particular cell is covered by the AR. It is illustrated by the Fig. 4. In Fig. 4, the initial position of the mobile user lies at $x = 0$. The user is moving with a velocity $v$. The area of communication in which the user is moving is covered by the AR such that the intruder attacks on the mobile user at $x = d$. However, to prevent or to reduce the impact of AR on the intruder, it is present at the boundary of the area covered by the AR. The attack is initiated by the intruder on the target user either at the ($t = 0 + +, x = 0 + +$) or less than ($t = T, x = d$). It is due the fact that the intensity of the AR is moderate or less than the predefined value of the AR at ($t = 0 + +, x = 0$) or at ($t = T, x = d$), $d \leq 0.5 kms$. Therefore, even in case of mobile user the AR impact will be observed in the cell coverage area. Also, UAVs (Unmanned Aerial Vehicles) can prove an effective choice to attain the precision of the AR on a specific area.

## IV. HALF DUPLEX ATTACK: AN ILLUSTRATIVE EXAMPLE

This section demonstrates the equipped strategy of the proposal. Fig.5. depicts the comparative performance analysis in presence and in the absence of an intruder. Real-Time random deployment of the users is considered in view of cybersecurity wherein setup of FD and HD attacks are observed. Based on the channel modeling, channel gain assigned to different users under the inspection of the path loss model [16]. The distance of the user from the BS is denoted at the top of each bar.

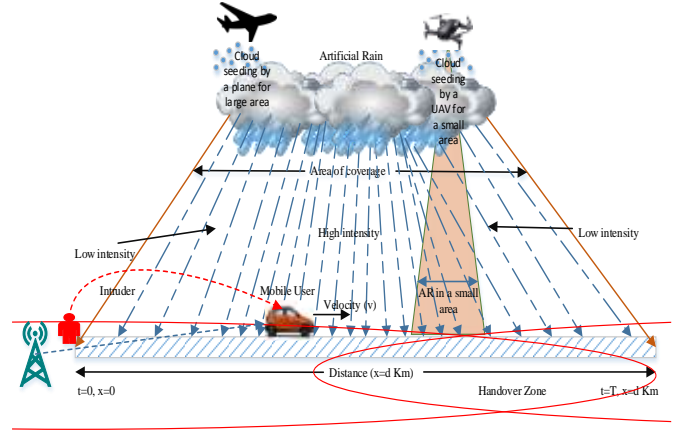

Fig. 4. Impact of AR on mobile user

*A. Intruder attack on User-3 in an urban scenario*

From the Fig.5(a), it is observed that the users are deployed under the random scenario, in which user 3 occupies the farthest distance (223m) from the BS and thus experiences maximum path loss as depicted in equation (15). More the distance of the user from the BS more is the path loss. Therefore, the user-3 occupies minimum gain associated with the BS. Thus, it is observed as the prime target for the attack. The intruder follows the scanning procedure for spoofing by the intruder. The performance of the user-3 is degraded by the presence of the AR. Under the implementation of the AR, it is perceived that chance of intruding the attack on the user 3 increases. Also, from Fig. 5(c), the degraded performance of the user-3 due to AR can be compared with the achieved performance of the user-2 of a similar distance.

*B. Intruder sensitivity in FD and HD in an urban scenario*

From Fig. 5(a) and 5(b), under the phenomenon of HD and FD spoofing attack in urban, HD attack is detected to be more sensitive in comparison to the FD attack. According to Wyner's theory of wiretap channels, the security of information is attained if the quality of the main channel is greater than the quality of the wiretap channel. Following the phenomenon of AR, the quality of the channel of user-3 is degraded, which enhances the chance of the intruder to spoof successfully. Therefore, the optimal quality for the wiretap channel, required for the attack which is given as $C_{ev} > C_u$. Thus the optimal value to spoof the resources from the valid user is the just higher value than $C_u$. The more optimal value is attained by HD attack, just above the throughput of the user-3, enables more sensitivity as compared to the sensitivity provided by the FD attack.

*C. Impact of AR in the urban and rural scenario*

From Fig. 5(b) and 5(d), the observed throughput under the execution of attack in the presence of AR, is less in the urban scenario than in the rural scenario. In other words, it depicts that the impact of AR is more significant in the urban scenario than in rural scenarios. Therefore, it can be concluded that the urban scenario is more prone to security attacks as compared to the rural scenario by using AR.

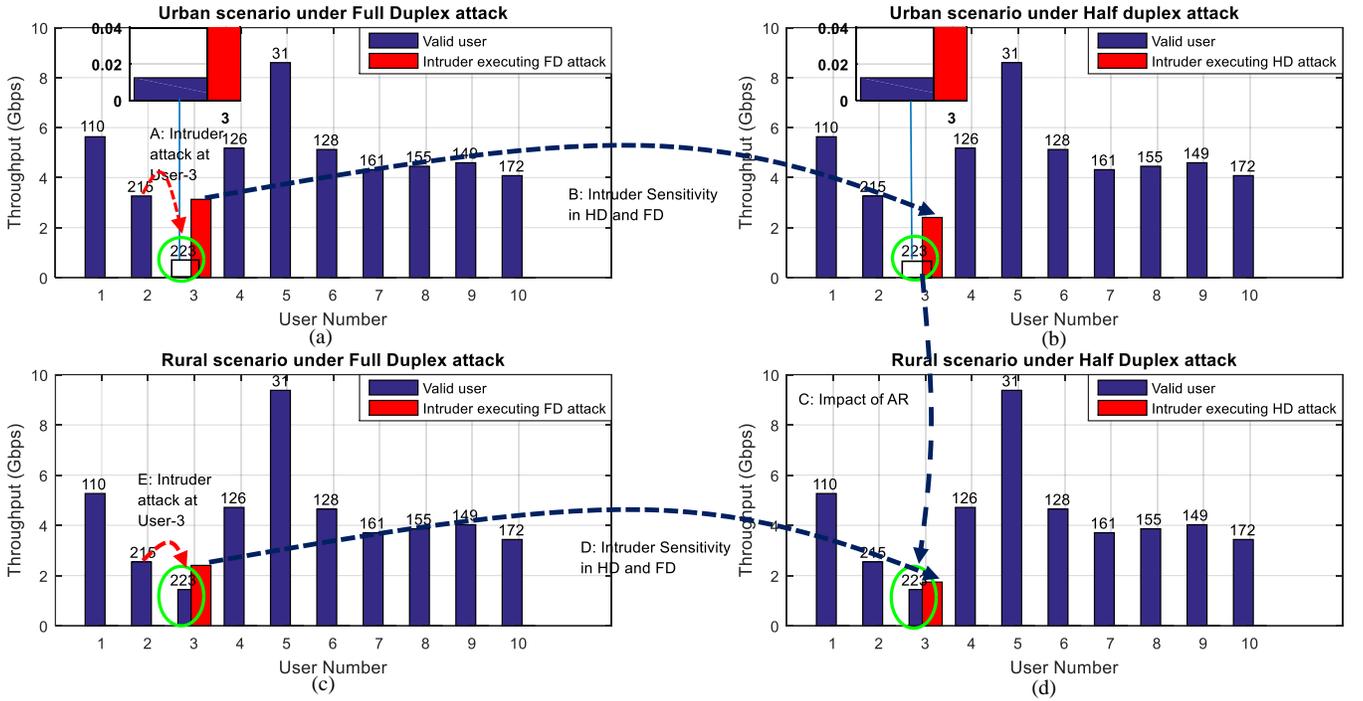

Fig. 5. General analysis of HD (Half Duplex) attack and FD (Full Duplex) attack

*D. Intruder sensitivity in FD and HD in an urban scenario*

A similar justification is followed in the rural scenario as given for urban scenarios. However, the observed impact of the attack is comparatively less significant in the rural scenario than in urban scenarios at the user-3 deployed at the same distance from BS.

*E. Intruder attack on User-3 in the rural scenario*

From Fig. 5(c) the user-3 present at the extreme end of the cell at a distance 223m and the user-2 present at a similar distance of 215m from BS. Though similar channel conditions are expected to be obtained for both the users, nonetheless, the impact of AR and attack decreases the throughput of user-3. Comparing the rural and urban scenarios, the rural scenario is concluded to be more optimal as far as the security of the network is concerned in the presence of AR.

V. RESULTS AND DISCUSSION

In this section, the results of the simulation are examined and discussed systematically. For investigating the results, urban scenarios and rural scenarios under the microcell deployment are taken into the assessment. The security analysis is evaluated based on the performance of the valid user in the presence of an intruder executing an HD attack. This section is categorized into three main segments. The setup analysis is specified in the first segment. The second segment examines the security analysis and performance analysis in the presence of AR. The last segment summarizes the analysis in brief.

*A. Simulation setup*

The scenario of the proposed schematic and the analysis on different parameters are carried out in MATLAB computing environment. Table-IV defines the simulation parameters used for the evaluation.

*B. Result analysis*

A comparative analysis is made between the rural and urban scenarios in the presence and absence of the artificial rain while executing HD attack and FD attack. Fig. 6. demonstrates secrecy rate analysis of the valid user with respect to the distance in rural and urban scenarios under the execution of AR at an operating frequency of 28 GHz. It is illustrated in the figure, the secrecy rate achieved in urban scenarios under the influence of AR decreases rapidly with an approximation of 4.94 GHz. Also, in the case of a rural scenario, the overall difference of the secrecy rate in the presence of the AR is approximated as 0.95 Gbps. It defines that the impact of AR is thoroughly observed in an urban scenario than in a rural scenario. However, the relative difference of the secrecy rate in the presence of AR is identified as 4.29 Gbps in the urban scenario and 9.48 Gbps in the rural scenario from the distance of 10m to 250m. In addition to it, the relative difference in the secrecy rate of rural scenario is 8.58 Gbps, and in the urban scenario is of the range of 6.58 Gbps from the distance of 10m to 250m in the absence of AR, which is comparatively lower in urban than in rural scenario. Therefore, from the above analysis, it is concluded that the urban scenario executes a drastic decrease in secrecy rate than in rural scenario for a distance of 10m to 250m in the presence of AR.

Fig. 7. shows the energy efficiency of the user at different distances from the BS. A steep decreasing slope is observed for

TABLE IV
SIMULATION PARAMETERS

| Parameter | Value | Parameter | Value |
|---|---|---|---|
| Bandwidth | 800MHz | Frequency | 28GHz |
| Noise power | -106dBm | Rain rate | 50mm/hr |
| Transmission power | 20 mW | Coverage range | 250m |

the urban scenario, with a huge difference of 135.39 Gbps decrease due to the presence of AR. For the rural scenario, close proximity is observed between the presence of the effect of AR and in the absence of the AR with a factor of 47.97 Gbps in contrary to the urban scenario.

From Fig. 8. secrecy rate analysis is witnessed under the impact of AR for different frequencies, including mmWave frequencies. The results are performed under the distance of 100m of the user from the BS. An exponential decrease of secrecy rate is observed with an increase in the frequency. For rural and urban scenarios, rural scenario occupies a more secrecy rate as compared to the secrecy rate achieved in an urban scenario with respect to the increase in frequency. In the presence of AR, the rural scenario is observed to have a drastic impact of AR on the achieved secrecy rate than the secrecy rate achieved in the case of an urban scenario. From the above analysis, it is evaluated that with an increase in frequency, the secrecy rate tends to decrease exponentially, and in the presence of AR, the secrecy rate in the rural scenario has a more significant impact.

Fig. 9. shows the energy efficiency for millimeter-wave frequencies in the urban and rural scenarios. A decreasing exponential characteristic of the energy efficiency is depicted for the urban scenario for an increase in frequency. However, better performance of energy efficiency is observed in the rural scenario than in urban scenarios. Under the analysis of AR, a rural scenario provides relatively less difference in the performance of energy efficiency than energy efficiency achieved in an urban scenario. In other words, the urban scenario has a greater impact of AR on the performance of energy efficiency with respect to the frequency.

Fig. 10 represents a miss rate for HD and FD attack under AR impact with respect to the number of attempts. It is determined from the graphical analysis, that a linearly increasing characteristic is shown between the number of attempts and the miss-rate. However, the proposed approach of the HD attack specifies lower miss-rate than the FD attack. The graph also represents the detail, with an increase in several attempts, the difference between miss-rate of HD and FD progressively increases.

Fig. 11. signifies the probability of missed attempts with respect to the number of exactly occurring missed attempts. The probabilities are obtained based on the Binomial (special case-Bernoulli) distribution and are compared with the probabilities achieved from the Poisons discrete distribution. The simulation is performed under AR in the communication network. The graph incorporates bell-curve characteristics in which HD attack possesses a lower probability of the miss-rate attempts as compared to the probability of miss rate attempts obtained in the FD attack. It is apparent from the graph that Binomial distribution provides a higher probability of missed attempts than the probability obtained from the distribution of the poison. Therefore, it can be concluded that the HD attack provides fewer chances of miss-attempts of attack. The obtained result motivates the enactment of the proposed

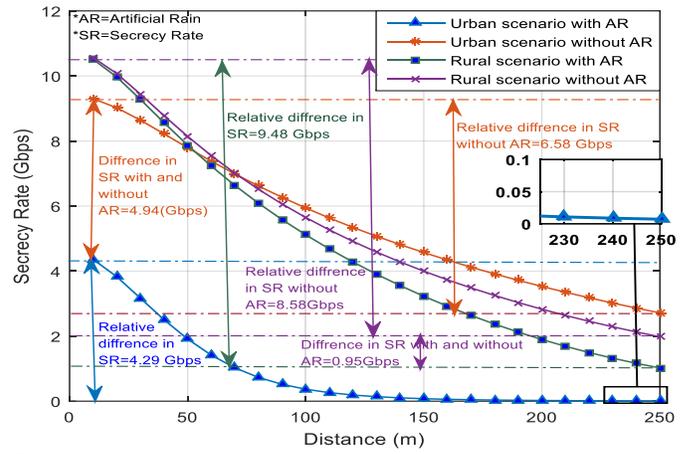
Fig. 6. Secrecy rate versus distance in rural and urban scenario in the presence and absence of artificial rain

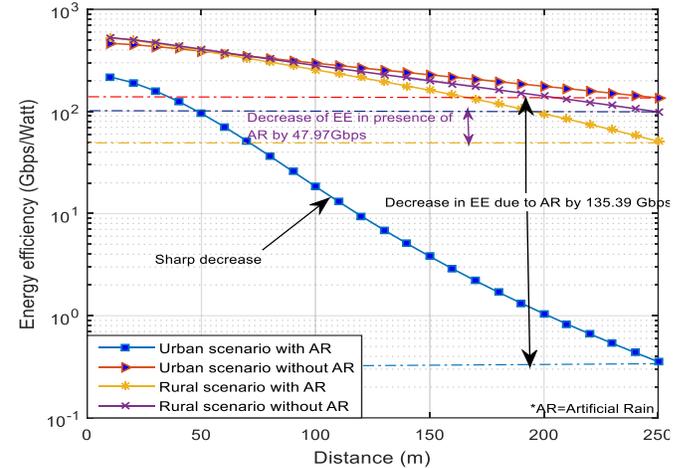
Fig. 7. Energy efficiency versus distance in rural and urban scenarios in the presence and absence of artificial rain

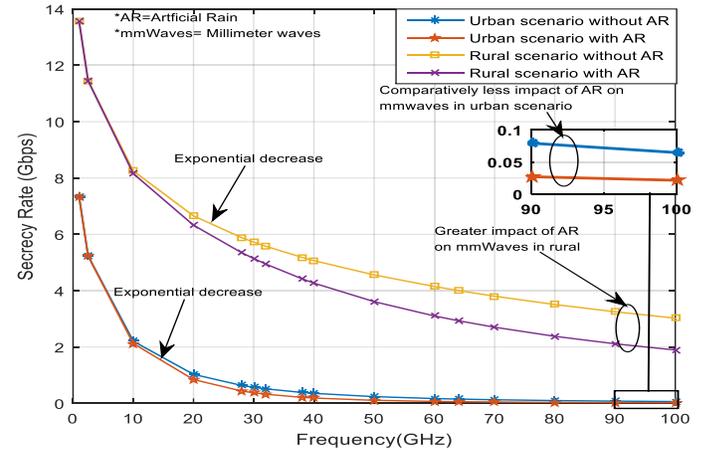
Fig. 8. Secrecy rate versus frequency in rural and urban scenario in the presence and absence of artificial rain

approach of HD attack for the successful execution of intruding.

Fig. 12. depicts the sensitivity of the proposed HD attack and the FD attack. The parameter of sensitivity defines the ability to sense the minimum capacity required to execute the successful existence of the attack on the valid user. The analysis is estimated under the farthest cell boundary conditions of the user from the BS. The distance of the user

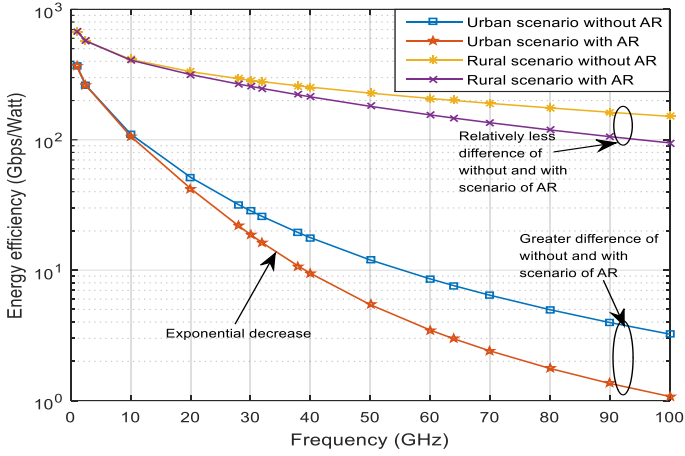

Fig. 9. Energy efficiency versus frequency in rural and urban scenarios in the presence and absence of AR.

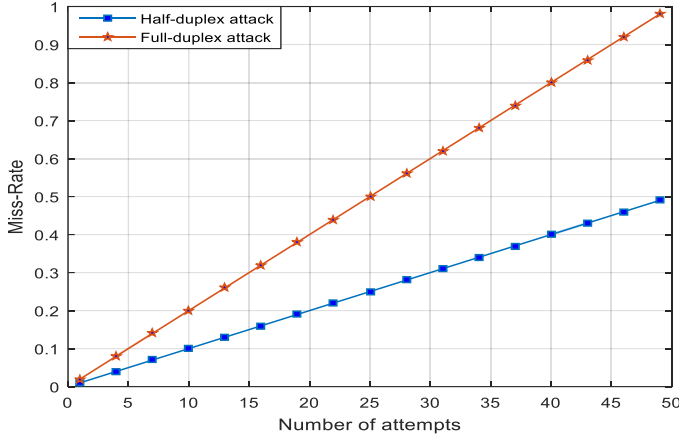

Fig. 10. Miss-rate versus the number of attempts

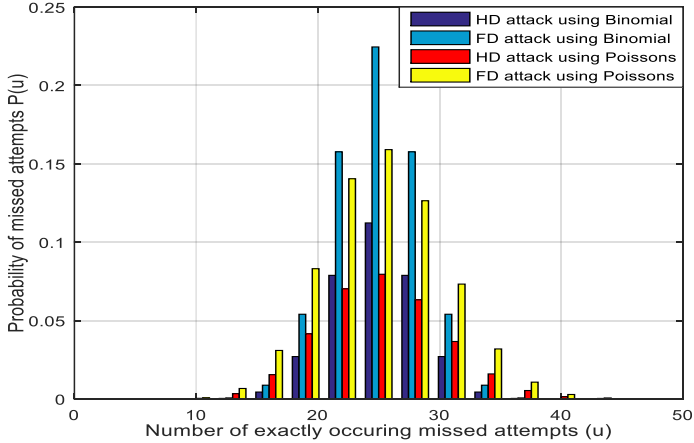

Fig. 11. Probability of missed attempts versus a number of exactly occurring missed attempts from a total number of attempts

from the BS is denoted at the top of each bar, respectively. The assessed results show that more distances of the user from the BS decreases the sensitivity, and thus decreases the probability of attaining an optimum value of capacity required by the attacker to decrease the secrecy capacity of the valid user below the threshold. Furthermore, it is also reflected in Fig. 12 that HD attack provides more sensitivity than an FD attack. The variation of sensitivity in FD and HD is considerably higher in the rural scenario than the variation of sensitivity for HD and FD attack in an urban scenario.

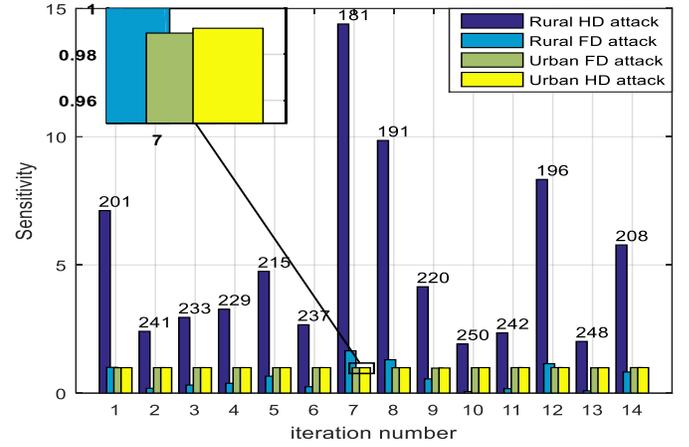

Fig. 12. Sensitivity analysis of the proposed attack and full-duplex for a randomly distributed distance of the user from the BS

*C. Summary of the observations and restrictions*

The HD based attack is designed for detection of the possible security menace to the users, significantly operated under the impact of AR to determine the security of the users in the rural and urban scenario. The executed approach simultaneously analyses the influence of AR, followed by the HD attack. The purpose of introducing AR is to worsen the secrecy rate of the valid user. At a particular distance, it is observed that the urban scenario has a greater impact of AR. However, under the relative difference of the secrecy rate for a range of distance 10m to 250m from the BS, the rural scenario undergoes greater deterioration in the secrecy rate. The observed results confirm that the performance of the HD attack is several times effective than that of the FD attack via the examination of the miss-rate. As a significance of which, the attacking capability of the proposed attack poses a greater threat. Moreover, under the framework of rural and urban scenario, the performance of the attack in a rural scenario is proved to be more efficient in terms of sensitivity, and energy efficiency.

Additionally, the proposed framework occupies a minimum radiation aspect with a reachable resilient signal strength for a distance of 120m from BS. However, for a macrocell radius greater than 250m, the transmission power is required to be approximated to 23dbm to 43dbm, though it results in an upsurge of interference and complexity.

## VI. CONCLUSION AND FUTURE WORK

In this paper, the security perspective of the rain attenuation is taken into consideration for the wireless communication network. The prime focus is the implementation of the HD attack model with better performance as compared to the conventional attacks for both rural and urban scenarios. The framework of the attack broadly consists of two parts. The first part defines the execution of the AR by the attacker to decrease the secrecy rate of the valid user. The second part involves the implementation of an HD attack. The parameter of distance is taken into concern for the evaluation of the security parameters. Furthermore, an analysis is performed on the mmWave frequencies in the rural and urban scenarios under the influence of AR. The attained results have emphasized that the proposed approach of HD attack proves to be more effectual in terms of miss-rate, and sensitivity.

In future work, the accuracy of the proposed attack can be increased by optimizing the required capacity to execute the attack. Also, proficient mechanisms of AR can be initiated to improve the cost efficiency and accuracy. Additionally, the proposed approach can be extended for testing of UAV (Unmanned Ariel Vehicle) application and can prove advantageous in border areas of defense to spoof the information from the opponent.

## APPENDIX

*Lemma 1:* The secrecy capacity of the user is reduced due to the AR attenuation introduced by the intruder, and the corresponding SNR gets deteriorated. The sum of secrecy capacity required to decrease the secrecy capacity below threshold capacity can be obtained by AR attenuation given by:

$$(C_s)_{AR} = C_u - (C_{ev})_{AR} < C_T \tag{39}$$

Thus, $$C_s > (C_s)_{AR} \tag{40}$$

*Proof:* Consider a scenario in which path loss $P_{l\ (db)}$ is obtained in the absence of an intruder given by:

$$P_{l\ (db)} = P_{t\ (db)} - P_{r\ (db)} \tag{41}$$

SNR can also be obtained as:

$$(\varsigma)_{dB} = P_{t\ (db)} - P_{l\ (db)} - \gamma' \tag{42}$$

The secrecy capacity is given by:

$$C_s = C_u - C_{ev} \tag{43}$$

In the absence of eavesdropper $C_{ev} \approx 0$

$$C_s = C_u$$
$$C_s = \log_2(1 + \varsigma) \tag{44}$$

For the scenario in the presence of an intruder incorporating AR. Let $\varpi$ be the attenuation created by AR. The path loss for this scenario is given by:

$$P'_{l\ (db)} = P_{t\ (db)} - P_{r\ (db)} - (\varpi)_{AR} \tag{45}$$

The equation of SNR can be expressed as:

$$(\varsigma^{AR})_{dB} = P_{t\ (db)} - P_{l\ (db)} - (\varpi)_{AR} - \gamma' \tag{46}$$

Comparing equation (42) and equation (46), we get:

$$(\varsigma^{AR})_{dB} = (\varsigma)_{dB} - (\varpi)_{AR} \tag{47}$$

Equation (47) indicates $(\varsigma)_{dB} > (\varsigma^{AR})_{dB}$. From equation (12) the capacity of the user is proportional to the SNR. Therefore, the corresponding secrecy capacities of SNR obtained in the presence of eavesdropper under the procedure of AR attenuation is given by:

$$(C_s)_{AR} = C_u - (C_{ev})_{AR} \tag{48}$$

Substituting the value of $C_u$ from equation (44) in equation (48), we get:

$$(C_s)_{AR} = C_s - (C_{ev})_{AR} \tag{49}$$

From equation (49), it is observed that $(C_s)_{AR}$ is a smaller quantity than $C_s$, therefore, we conclude that

$$(C_s)_{AR} < C_s \tag{50}$$

Let the amount of attenuation is created in such a manner such that the secrecy capacity reduces below the threshold given by:

$$(C_s)_{AR} = C_s - (C_{ev})_{AR} < C_T \tag{51}$$

Or $$(C_s)_{AR} = C_u - (C_{ev})_{AR} < C_T \tag{52}$$

Equation (52) proves the proposed lemma

The expression obtained from equation (52) indicates that the achieved secrecy capacity can be decreased by the attenuation of AR. Thereby, it can act as the performance limiting parameter. The illustration for the obtained results based on equation (52) is given in Fig. 6. and Fig.8.

*Lemma 2:* The probability of a miss rate for the HD attack is less than the probability of a miss rate for the FD attack. However, the miss rate probability can be equal for HD attack and FD attack for zero miss rate probability in the uplink

$$P(m_r)_{HD} \leq P(m_r)_{FD} \tag{53}$$

*Proof:* Let $J$ and $K$ define the miss rates for uplink and downlink, respectively. The HD attack involves the attack on downlink only. Therefore, the miss rate probability for the half-duplex attack is given by:

$$P(m_r)_{HD} = P(K) \tag{54}$$

The miss rate probability for FD attack can be given as:

$$P(m_r)_{FD} = P(J + K) \tag{55}$$
$$= P(J) + P(K) - P(J \cap K) \tag{56}$$

The events $J$ and $K$ are mutually exclusive and are I.I.D. therefore, $$P(J \cap K) = 0 \tag{57}$$

Equation (56) becomes

$$P(m_r)_{FD} = P(J) + P(K) \tag{58}$$

Using equation (54)

$$P(m_r)_{FD} = P(J) + P(m_r)_{HD} \tag{59}$$

For minimum miss rate probability in the uplink, such that $P(J) \approx 0$ equation (59) can be given as:

$$P(m_r)_{FD} = P(m_r)_{HD} \quad \forall \ P(J) \approx 0 \tag{60}$$

However, probability cannot be negative, for miss rate probability in uplink greater than zero. Equation (59) can be obtained as:

$$P(m_r)_{FD} > P(m_r)_{HD} \quad \forall \ P(J) > 0 \tag{61}$$

Combining equation (60) and (61) we get:

$$P(m_r)_{FD} \geq P(m_r)_{HD} \quad \forall \ P(J) \geq 0$$

Or $$P(m_r)_{HD} \leq P(m_r)_{FD} \ \forall \ P(J) \geq 0 \tag{62}$$

Equation (62) proves the stated lemma and derives the condition that the probability of the miss rate of the attacking attempts for HD attack cannot be greater than the FD attack. The illustration of the achieved expression is depicted in Fig. 11.

*Corollary 1:* The miss-rate achieved for the HD attack $(m_r)_{HD}$ on device-1 by an intruder is less than or equal to miss rate of intruder achieved in the FD attack $(m_r)_{FD}$.

$$(m_r)_{HD} \leq (m_r)_{FD}$$

*Proof:* Consider $x$ as the total number of attempts made by the intruder and $m$ number of missed attempts to access resource allocation in HD attack such that miss rate for HD attack can be expressed as:

$$(m_r)_{HD} = \frac{m}{x} \tag{63}$$

For FD attack, let $m_1$ and $m_2$ denotes the number of missed attempts in uplink and downlink, respectively. The miss rate for FD can be given as:

$$(m_r)_{FD} = \frac{m_1+m_2}{x} \quad (64)$$

$$(m_r)_{FD} = \frac{m_1}{x} + \frac{m_2}{x} \quad (65)$$

For HD attack, the number of missed attempts are defined for downlink only. Let the number of missed attempts in HD be an equal number of missed attempts in the downlink for the FD attack. Equation (64) can be re-written as:

$$(m_r)_{FD} = \frac{m_1}{x} + \frac{m}{x} \quad (66)$$

Comparing equation (63) and equation (66), we get

$$(m_r)_{FD} = (m_r)_{HD} + \frac{m_1}{x} \quad (67)$$

Where $\frac{m_1}{x}$ is non-negative quantity and lies between 0 to $n$. Therefore,

$$(m_r)_{HD} \leq (m_r)_{FD} \quad \forall \; \frac{m_1}{x} \geq 0 \quad (68)$$

Equation (68) proves the miss rate in HD attack is less than the miss-rate observed in the FD attack for an equal number of missed attempts in respective downlinks. However, for zero missed attempts in the uplink of the FD attack, the miss rate for the FD attack, in that case, is equal to the miss rate in HD attack.

## ACKNOWLEDGMENT

The authors gratefully acknowledge the support provided by 5G and IoT Lab, DoECE, and TBIC, Shri Mata Vaishno Devi University, Katra, Jammu and Kashmir, India.

## REFERENCES

[1] David, K., & Berndt, H. (2018). 6G vision and requirements: Is there any need for beyond 5G?. *IEEE vehicular technology magazine*, *13*(3), 72-80.

[2] Dai, L., Wang, B., Ding, Z., Wang, Z., Chen, S., & Hanzo, L. (2018). A survey of non-orthogonal multiple access for 5G. *IEEE Communications Surveys & Tutorials*, *20*(3), 2294-2323.

[3] Abrol, A., & Jha, R. K. (2016). Power optimization in 5G networks: A step towards GrEEn communication. *IEEE Access*, *4*, 1355-1374.

[4] Parvez, I., Rahmati, A., Guvenc, I., Sarwat, A. I., & Dai, H. (2018). A survey on low latency towards 5G: RAN, core network, and caching solutions. *IEEE Communications Surveys & Tutorials*, *20*(4), 3098-3130.

[5] Gupta, A., Jha, R. K., & Jain, S. (2017). Attack modeling and intrusion detection system for 5G wireless communication network. *International Journal of Communication Systems*, *30*(10), e3237.

[6] Devi, R., Jha, R. K., Gupta, A., Jain, S., & Kumar, P. (2017). Implementation of intrusion detection system using adaptive neuro-fuzzy inference system for 5G wireless communication network. *AEU-International Journal of Electronics and Communications*, *74*, 94-106.

[7] Zhang, Y. P., Wang, P., & Goldsmith, A. (2015). Rainfall effect on the performance of millimeter-wave MIMO systems. *IEEE Transactions on wireless communications*, *14*(9), 4857-4866.

[8] Luini, L., Emiliani, L., Boulanger, X., Riva, C., & Jeannin, N. (2017). Rainfall rate prediction for propagation applications: model performance at regional level over Ireland. *IEEE Transactions on Antennas and Propagation*, *65*(11), 6185-6189.

[9] Kamga, G. N., & Aïssa, S. (2018). Wireless power transfer in mmWave massive MIMO systems with/without rain attenuation. *IEEE Transactions on Communications*, *67*(1), 176-189.

[10] Singh, R., & Acharya, R. (2018). Development of a New Global Model for Estimating One-Minute Rainfall Rate. *IEEE Transactions on Geoscience and Remote Sensing*, (99), 1-7.

[11] Arslan, C. H., Aydin, K., Urbina, J. V., & Dyrud, L. (2017). Satellite-link attenuation measurement technique for estimating rainfall accumulation. *IEEE Transactions on Geoscience and Remote Sensing*, *56*(2), 681-693.

[12] Xu, J., Chen, L., Liu, K., & Shen, C. (2018). Designing Security-Aware Incentives for Computation Offloading via Device-to-Device Communication. *IEEE Transactions on Wireless Communications*, *17*(9), 6053-6066.

[13] Wang, J., Huang, Y., Jin, S., Schober, R., You, X., & Zhao, C. (2018). Resource management for device-to-device communication: A physical layer security perspective. *IEEE Journal on Selected Areas in Communications*, *36*(4), 946-960.

[14] Wang, L., Liu, J., Chen, M., Gui, G., & Sari, H. (2018). Optimization-based access assignment scheme for physical-layer security in D2D communications underlaying a cellular network. *IEEE Transactions on Vehicular Technology*, *67*(7), 5766-5777.

[15] Sun, L., Du, Q., Ren, P., & Wang, Y. (2015). Two birds with one stone: Towards secure and interference-free D2D transmissions via constellation rotation. *IEEE Transactions on Vehicular Technology*, *65*(10), 8767-8774.

[16] 3GPP Technical Specifications TR 38.900 (Release 15) 5G group radio access network, June 2018.

[17] Parkvall, S., Dahlman, E., Furuskar, A., & Frenne, M. (2017). NR: The new 5G radio access technology. *IEEE Communications Standards Magazine*, *1*(4), 24-30.

[18] Wallace, J. M., & Hobbs, P. V. (2006). *Atmospheric science: an introductory survey* (Vol. 92). Elsevier.

[19] North, G. R., Pyle, J. A., & Zhang, F. (Eds.). (2014). Encyclopedia of atmospheric sciences (Vol. 1). Elsevier.

[20] Doshi, N., & Agashe, S. (2015). Feasibility study of artificial rainfall system using ion seeding with high voltage source. *Journal of Electrostatics*, *74*, 115-127.

[21] G. Chopra, R. K. Jha and S. Jain, "TPA: Prediction of Spoofing Attack Using Thermal Pattern Analysis in Ultra Dense Network for High Speed Handover Scenario," in *IEEE Access*, vol. 6, pp. 66268-66284, 2018.

[22] A. Gupta, R. K. Jha, P. Gandotra and S. Jain, "Bandwidth Spoofing and Intrusion Detection System for Multistage 5G Wireless Communication Network," in *IEEE Transactions on Vehicular Technology*, vol. 67, no. 1, pp. 618-632, Jan. 2018.

[23] S. Parkvall, E. Dahlman, A. Furuskar and M. Frenne, "NR: The New 5G Radio Access Technology," in *IEEE Communications Standards Magazine*, vol. 1, no. 4, pp. 24-30, Dec. 2017.

[24] 5G NR-Driving Wireless Evolution into New Vertical Domains "Intel Next Generation and Standards," Aug. 2018 https://www.intel.com/content/dam/www/public/us/en/documents/guides/5g-nr-technology-guide.pdf

[25] Scientific and technical Aerospace Reports "*National Aeronautics and Space Administration (NASA)*"vol. 8, no. 7, April, 1970

[26] S. Han, Y. Zhang, W. Meng, C. Li and Z. Zhang, "Full-Duplex Relay-Assisted Macrocell with Millimeter Wave Backhauls: Framework and Prospects," in *IEEE Network*, vol. 33, no. 5, pp. 190-197, Sept.-Oct. 2019.

[27] S. Han, Y. Zhang, W. Meng and H. Chen, "Self-Interference-Cancelation-Based SLNR Precoding Design for Full-Duplex Relay-Assisted System," in *IEEE Transactions on Vehicular Technology*, vol. 67, no. 9, pp. 8249-8262, Sept. 2018.

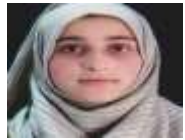
**Misbah Shafi** is currently pursuing a Ph.D. degree in electronics and communication engineering at SMVD University, J&K, India. Currently, she is doing her research on security issues in next generation communication networks.

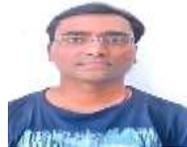
**Dr. Rakesh K Jha (S'10, M'13, SM 2015)** is currently an Associate Professor in the School of Electronics and Communication Engineering, SMVD University, Katra, Jammu and Kashmir, India. He has published more than 41 Science Citation Index Journals Paper with more than 1941 Citations in his credit. He has attended 25 International Conference papers. His area of interest is Wireless communication, Optical Fiber Communication, Computer Networks, and Security issues. Dr. Jha's one concept related to the router of Wireless Communication was accepted by ITU (International Telecommunication Union) in 2010and.

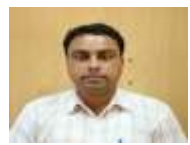
**Dr. Manish Sabraj** is currently an Assistant Professor in the School of Electronics and Communication Engineering, SMVD University, Katra, J&K, India. His research interest includes Wireless Sensor Networks (WSN), and Spectrum analysis.